\documentstyle[aps,multicol,fleqn,epsf,psfig]{revtex}

\newcommand{\beq}{\begin{equation}}
\newcommand{\eeq}{\end{equation}}
\newcommand{\beqa}{\begin{eqnarray}}
\newcommand{\eeqa}{\end{eqnarray}}

\begin{document}

\title{Quantum Search by Local Adiabatic Evolution}
\author{J\'er\'emie Roland$^1$ and Nicolas J. Cerf$^{1,2}$}
\address{$^1$ Ecole Polytechnique, CP 165, 
Universit\'e Libre de Bruxelles, 1050 Brussels, Belgium\\
$^2$ Jet Propulsion Laboratory, California Institute of Technology, 
Pasadena, California 91109\\}

\date{July 2001}
\draft

\maketitle

\begin{abstract}
The adiabatic theorem has been recently used to design quantum algorithms 
of a new kind, where the quantum computer evolves slowly enough 
so that it remains near its instantaneous ground state 
which tends to the solution\cite{farhi1}.
We apply this time-dependent Hamiltonian approach to the Grover's problem,
i.~e., searching a marked item in an unstructured database. We find that, 
by adjusting the evolution rate of the Hamiltonian so as to keep
the evolution adiabatic on each infinitesimal time interval,
the total running time is of order $\sqrt{N}$, where $N$ is the number
of items in the database. We thus recover the advantage of Grover's
standard algorithm as compared to a classical search, scaling as $N$.
This is in contrast with the constant-rate adiabatic approach
developed in \cite{farhi1}, where the requirement of adiabaticity 
is expressed only globally, resulting in a time of order $N$.
\end{abstract}


\begin{multicols}{2}

\subsection*{Introduction}

Although quantum computation is mostly a theoretical concept today,
several quantum algorithms have already been designed and shown to
outperform all known classical algorithms, thereby giving a strong
motivation to the development of quantum computers.  Probably the most
spectacular result is Shor's algorithm\cite{shor}, which can factor a large
number with a computation time polynomial in the size of the number,
whereas all known classical algorithms require a (sub-)exponential
time. Another remarkable algorithm, due to Grover, concerns the problem
of searching in an unsorted database\cite{grover}. Suppose we have a database 
of $N$ items, one of which is marked. The goal is to find this unknown 
marked item by accessing the database a minimum number of times. 
Classically, $N/2$ items must be tested, on average, before finding 
the right one. Grover's quantum algorithm performs the same task with
a complexity of order $\sqrt{N}$, giving rise to a quadratic speed up.

While Grover's algorithm was originally presented within the 
standard paradigm for quantum computation, that is using a discrete 
sequence of unitary logic gates, we will turn here to
another type of quantum computation where the state of the quantum
register evolves continuously under the influence of some driving
Hamiltonian. This concept of quantum computation viewed as
a continuous time evolution was pioneered by 
E. Farhi and S. Gutmann in \cite{farhi2}, where they proposed
an ``analog'' version of Grover's algorithm based on 
a time-independent Hamiltonian applied for a time $T$.   
Their algorithm required $T$ to be of order $\sqrt{N}$, which gives
thus the same complexity as Grover's algorithm. In a more recent article
with other coworkers, they considered an alternative class of continuous-time 
quantum algorithms based on a time-dependent Hamiltonian
that effects an {\em adiabatic} evolution of the quantum register\cite{farhi1}.
If the evolution of a quantum system is governed by a Hamiltonian 
that varies slowly enough, this system will stay 
near its instantaneous ground state. This adiabatic evolution can be
used to switch gradually from an initial Hamiltonian, 
whose ground state is known, to a final Hamiltonian, whose ground
state encodes the unknown solution. The time required for this
switching to remain globally adiabatic determines the computation time. 
Farhi {\it et al.} have solved Grover's search problem using 
this adiabatic evolution approach, 
but this unfortunately resulted in a complexity of order $N$,
that is no better than a classical algorithm that checks all possible
solutions\cite{farhi1}.

In the present article, we show that one can recover the quadratic speed-up
of Grover's original algorithm by continuously adjusting the rate with which
the initial Hamiltonian is switched to the final Hamiltonian
so as to fulfill the condition of adiabaticity {\em locally}, 
i.~e. at each time. Interestingly, this local adiabatic evolution approach 
makes it thus possible to improve the scaling law of the complexity 
of the quantum search algorithm simply by varying the evolution rate of
the driving Hamiltonian.
This offers the perspective of speeding up more sophisticated
adiabatic evolution algorithms, such as those applied to NP-complete
problems\cite{farhi3}. It might also be used to design an adiabatic evolution
version of the nested quantum search technique 
proposed in \cite{cerf} for solving structured problems.

\subsection*{Adiabatic theorem}

Consider a quantum system in a state $|\psi(t)\rangle$, which
evolves according to the Schr\"{o}dinger equation
\beq
i \frac{d}{dt} |\psi(t)\rangle\,=\,H(t)|\psi(t)\rangle
\eeq
where $H(t)$ is the Hamiltonian of the system (we let $\hbar=1$). 
If this Hamiltonian is time-independent and the system is initially in
its ground state, then it will remain in this state. 
The Adiabatic Theorem (see e.~g. \cite{bransden}) 
expresses that if the Hamiltonian varies slowly enough, roughly speaking,
the state of the system will stay close to the instantaneous ground state
of the Hamiltonian at each time $t$. 
More specifically, let $|E_k;t\rangle$ be the eigenstates of $H(t)$,
satisfying
\beq\label{eigenproblem}
H(t)|E_k;t\rangle\,=\,E_k(t)|E_k;t\rangle
\eeq
where $E_k(t)$ are the corresponding eigenvalues and
$k$ labels the eigenstates ($k=0$ labels the ground state).
We define the minimum gap between the lowest two eigenvalues as
\beq\label{g}
g_{\min}=\min_{0 \leq t \leq T} \left [ E_1(t)-E_0(t) \right ]
\eeq
and the matrix element of $dH/dt$ between the two corresponding eigenstates as
\beq\label{epsilon}
\langle \frac{dH}{dt} \rangle_{1,0}=\langle E_1;t|\frac {dH}{dt}|E_0;t\rangle
\eeq
The Adiabatic Theorem states that 
if we prepare the system at time $t=0$ in its ground state $|E_0;0\rangle$ 
and let it evolve under the Hamiltonian $H(t)$, then
\beq
|\langle E_0;T| \psi(T) \rangle|^2 \geq 1-\varepsilon^2
\eeq
provided that
\beq\label{adiabatic}
\frac{\left | \langle \frac{dH}{dt} \rangle_{1,0} \right |}{g^2_{\min}} 
\leq \varepsilon
\eeq
where $\varepsilon<<1$.
In particular, this implies that the minimum gap cannot be lower than
a certain value if we require the state at time $t$ to differ from the
instantaneous ground state by a negligible amount (a smaller gap
implies a higher transition probability to the first excited state).
This result can be used to design a new type 
of quantum algorithm based on a time-dependent Hamiltonian\cite{farhi1}. 
Assume we can build a Hamiltonian for which we know that the ground state
encodes the solution of a problem. Then, it suffices to prepare the system
in the ground state of another Hamiltonian, easy to build,
and change progressively this Hamiltonian into the other one in order to get, 
after measurement, the sought solution with large probability. 
The Adiabatic Theorem imposes the minimum time it
takes for this switching to be adiabatic, and this time 
can be thought of as the algorithm complexity.

\subsection*{Global vs local adiabatic evolution 
for solving the quantum search problem}

\begin{figure}
\begin{center}
\epsfxsize=8cm \epsffile{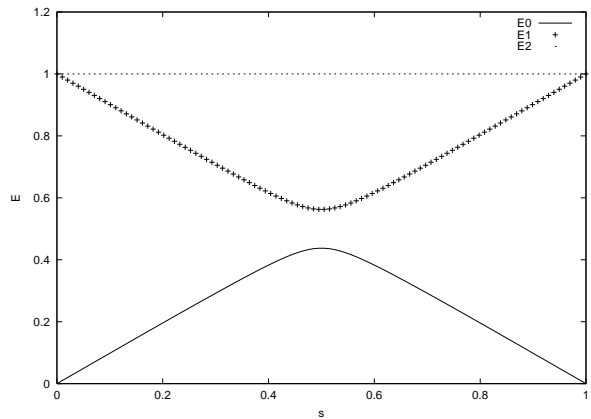}
\end{center}
\narrowtext
\caption{Eigenvalues of the time-dependent Hamiltonian $\tilde{H}(s)$ 
as a function of the reduced time $s$, for $N=64$.}
\label{graph1}
\end{figure}

We now apply this adiabatic evolution method 
to the problem of finding an item in an unsorted database.
Before describing our local adiabatic search algorithm, 
we first summarize the method of \cite{farhi1},
based on a global adiabatic evolution.
Consider a set of $N$ items among which one is marked, the goal 
being to find it in a minimum time. We use $n$ qubits to label the items,
so that the Hilbert space is of dimension $N=2^n$. 
In this space, the basis states are written 
$|i \rangle$, with $i=0,\dots,N-1$, while 
the marked state is denoted by $|m \rangle$.
As we do not know $|m\rangle$ a priori, 
we use as an initial state an equal superposition 
of all basis states:
\beq
| \psi_0 \rangle\,=\,\frac{1}{\sqrt{N}}\sum_{i=1}^{N-1}|i\rangle
\eeq
We define two Hamiltonians:
\beqa
H_0 &=& I-|\psi_0\rangle\langle\psi_0| \\
H_m &=& I-|m\rangle\langle m|
\eeqa
whose ground states are $|\psi_0\rangle$ and $|m\rangle$, respectively, 
each with eigenvalue $0$. The time-dependent Hamiltonian underlying
the algorithm is a linear interpolation between these Hamiltonians,
that is
\beq\label{hamiltonian-t}
H(t)=(1-t/T)\,H_0 + t/T\,H_m
\eeq
or, with $s=t/T$
\beq\label{hamiltonian}
\tilde{H}(s)= (1-s)\,H_0 + s\,H_m
\eeq
The algorithm consists in preparing the system in the state
$|\psi(0)\rangle=|\psi_0\rangle$ and then applying the
Hamiltonian $H(t)$ during a time $T$. First, we notice that
\beq
\langle \frac{dH}{dt}\rangle_{1,0} = \frac{ds}{dt}\, 
\langle \frac{d\tilde{H}}{ds} \rangle_{1,0}
= \frac{1}{T}\, \langle \frac{d\tilde{H}}{ds} \rangle_{1,0}
\eeq
Using Eq.~(\ref{hamiltonian-t}) or (\ref{hamiltonian}),
we can solve the eigenproblem (\ref{eigenproblem}) 
and evaluate (\ref{g}) and (\ref{epsilon}).
The eigenvalues of $\tilde{H}(s)$ are plotted as a function
of $s$ in Fig.~\ref{graph1}. The highest eigenvalue $E_2=1$ 
is $N-2$ times degenerated, while the two lowest ones $E_0$ and $E_1$ 
have a degeneracy one. The difference between these lowest two 
eigenvalues defines the gap $g$. We find
\beqa
& & g=\sqrt{1-4 \frac{N-1}{N}s\,(1-s)} \label{gs} \\
& & \left|\langle \frac{d\tilde{H}}{ds}\rangle_{1,0}\right|\leq 1 \label{dhds}
\eeqa

We see that the minimum gap $g_{min}=1/\sqrt{N}$ is attained for $s=1/2$,
so that the adiabatic condition
(\ref{adiabatic}) is verified provided that
\beq  \label{T-farhi}
T \geq \frac{N}{\varepsilon}
\eeq
Thus, the computation time is of order $N$, and there is no advantage 
of this method compared to a classical search.

\begin{figure}
\begin{center}
\epsfxsize=8cm \epsffile{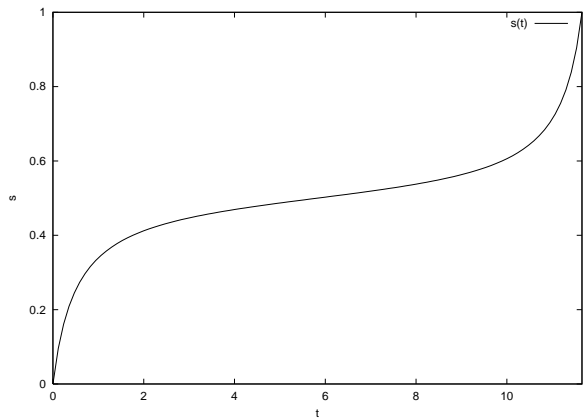}
\end{center}
\narrowtext
\caption{Dynamic evolution of the Hamiltonian
that drives the initial ground state to the solution state:
plot of the evolution function $s(t)$ for $N=64$. 
The global adiabatic evolution 
method of \protect\cite{farhi1}
would appear here as a straight line between $s(0)=0$ and $s(1)=1$.}
\label{graph2}
\end{figure}

Now, let us see how to improve on this adiabatic evolution method.
One should note that by applying Eq. (\ref{adiabatic}) globally, i.~e. 
to the entire time interval $T$, we impose a limit on the evolution rate
during the whole computation while this limit is only severe for the times 
around $s=1/2$, where the gap $g$ is minimum. Thus, by dividing $T$ into
infinitesimal time intervals $dt$ and applying the adiabaticity
condition locally to each of these intervals, we can vary the evolution
rate continuously in time, thereby speeding up the computation. In other
words, we do not use a linear evolution function $s(t)$ any more, but we adapt
the evolution rate $ds/dt$ to the local adiabaticity condition.
Let us find the optimum $s(t)$ with the boundary conditions 
$s(0)=0$ and $s(T)=1$.
Applying Eq. (\ref{adiabatic})
to each infinitesimal time interval, we get the new condition
\beq
\left|\frac{ds}{dt}\right| \leq 
\varepsilon \ \frac{g^2(t)}{\left| \langle \frac{d\tilde{H}}{ds} \rangle_{1,0}
\right|} 
\eeq
for all times $t$.
Using Eqs. (\ref{gs}) and (\ref{dhds}), we choose 
to make the Hamiltonian evolve at a rate that is solution of
\beq
\frac{ds}{dt}=\varepsilon \,g^2(t)=\varepsilon \left [1-4\frac{N-1}{N}s\,(1-s) \right ]
\eeq
where $\varepsilon<<1$. After integration, we find
\beqa   \label{t(s)}
t & = & \frac{1}{2\varepsilon}\,\frac{N}{\sqrt{N-1}} \left [ \arctan \left (\sqrt{N-1}(2s-1) \right ){} \right . \nonumber\\
& & {} \left . +\arctan \sqrt{N-1}  \right ]
\eeqa
By inverting this function, we obtain $s(t)$ as plotted in Fig.~\ref{graph2},
which shows the gradual change in
the switching between $H_0$ and $H_m$. 
We see that $H(t)$ is changing faster 
when the gap $g(s)$ is large while it evolves slower 
when $s$ is close to $1/2$, that is where the gap is minimum.
We may now evaluate the computation time of our new algorithm 
by taking $s=1$. With the approximation $N>>1$, we obtain
\beq\label{complexity}
T = \frac{\pi}{2 \varepsilon} \sqrt{N}
\eeq
improving upon Eq. (\ref{T-farhi}).
As a consequence,
we thus have a quadratic speed-up compared to a classical search,
and this algorithm can be viewed as the adiabatic evolution 
version of Grover's algorithm. In the Appendix, we show
that this algorithm is optimal, that is the computation time
cannot be shorter than $O(\sqrt{N})$ using any other 
evolution function $s(t)$.

\subsection*{Conclusion}

In this article, we have applied the adiabatic evolution technique
of \cite{farhi1} to design a quantum algorithm for solving
Grover's problem, i.~e., the search for a marked item 
in an unstructured database. We have shown
that applying the Adiabatic Theorem globally (as in \cite{farhi1}) imposes 
a running time of order $N$, where $N$ is the number of items in the database,
whereas adjusting the evolution rate of the Hamiltonian continuously 
in time so as to fulfill the adiabaticity condition locally
results in a time of order $\sqrt{N}$. We
therefore recover the advantage of Grover's usual algorithm
compared to a classical search \cite{grover}. We should notice that
this speed up was achieved by switching the
Hamiltonian according to Eq. (\ref{t(s)}),
which is only possible because here the gap $g(s)$ can
be derived analytically and does not depend on the solution 
of the problem (in our case, $|m\rangle$). As long as these conditions 
are satisfied, such a local adiabatic evolution method 
could be applied to more complicated
-- and more realistic -- problems such as NP-complete problems, 
treated using either a quantum
adiabatic evolution algorithm\cite{farhi3} or a nested version
of Grover's algorithm exploiting the problem structure\cite{cerf}.

J. R. acknowledges support from the Belgian FRIA.  N. C. is funded in part
by the project EQUIP under the IST-FET-QJPC European programme.

\subsection*{Appendix: Proof of optimality}

In this Appendix, we show that using our algorithm, 
no other choice of the evolution function $s(t)$ 
could lead to a better complexity than Eq. (\ref{complexity}).
This proof follows closely the lines
of the optimality proof of the ``analog'' Grover's algorithm
based on a time-independent Hamiltonian\cite{farhi2}.

Let $|\psi_m,t\rangle$ be the state of our quantum register during the
computation when the solution state is $|m\rangle$. 
After the computation time $T$, the states corresponding
to different solutions ($m$ and $m'$) must be sufficiently different:
\beq\label{condition}
1-|\langle\psi_m,T|\psi_{m'},T\rangle|^2\geq\varepsilon 
\qquad \forall m \neq m'
\eeq

Let us decompose the Hamiltonian $\tilde{H}(s)$ in two parts:
\beq
\tilde{H}(s)=\tilde{H}_1(s)+\tilde{H}_{2m}(s)
\eeq
where
\beqa
\tilde{H}_1(s)&=&I-(1-s) |\psi_0\rangle\langle \psi_0| \\
\tilde{H}_{2m}(s)&=&-s |m\rangle\langle m|
\eeqa
$|\psi_m\rangle$ and $|\psi_{m'}\rangle$ are solution of the Schr\"{o}dinger equations
\beqa
i\frac{d}{dt}|\psi_m,t\rangle&=&(H_1+H_{2m})|\psi_m,t\rangle \label{schro1}\\
i\frac{d}{dt}|\psi_{m'},t\rangle&=&(H_1+H_{2m'})|\psi_{m'},t\rangle \label{schro2}
\eeqa
with initial conditions
\beq\label{ci}
|\psi_m,0\rangle=|\psi_{m'},0\rangle=|\psi_0\rangle
\eeq
We will now derive a necessary condition on $T$ 
for Eq. (\ref{condition}) to be satisfied.
Using Eqs. (\ref{schro1}) and (\ref{schro2}), we have
\beqa
\hspace{-0.9 cm} &&\frac{d}{dt} [1-|\langle\psi_m,t|\psi_{m'},t\rangle|^2] \\
\hspace{-0.9 cm} &=&2\ \hbox{Im}\left [\langle \psi_m,t|H_{2m}-H_{2m'}|\psi_{m'},t\rangle \langle\psi_{m'},t|\psi_m,t\rangle \right ] \\
\hspace{-0.9 cm} &\leq&2 \ |\langle \psi_m,t|H_{2m}-H_{2m'}|\psi_{m'},t\rangle| \ |\langle\psi_{m'},t|\psi_m,t\rangle| \\
\hspace{-0.9 cm} &\leq&2\ \left [ |\langle \psi_m,t|H_{2m}|\psi_{m'},t\rangle| + |\langle \psi_m,t|H_{2m'}|\psi_{m'},t\rangle| \right ]
\eeqa
Summing over $m$ and $m'$, we get:
\beqa
&&\frac{d}{dt} \sum_{m,m'} [1-|\langle\psi_m,t|\psi_{m'},t\rangle|^2] \\
&\leq&4\sum_{m,m'}|\langle \psi_m,t|H_{2m}|\psi_{m'},t\rangle| \\
&\leq&4\sum_{m,m'} \|H_{2m}|\psi_m,t\rangle \| \, \| \, |\psi_{m'},t\rangle \| \\
&\leq&4N\sum_m \|H_{2m}|\psi_m,t\rangle \|
\eeqa
where we used the Cauchy-Schwartz inequality along with the fact that $|\psi_{m'},t\rangle$ is normalized.
The property
\beq
\hspace{-0.5 cm} \sum_m \| H_{2m}|\psi,t\rangle\|^2 = s^2 \Rightarrow \sum_m \| H_{2m}|\psi,t\rangle\|\leq \sqrt{N} s
\eeq
then leads to
\beq
\frac{d}{dt} \sum_{m,m'} [1-|\langle\psi_m,t|\psi_{m'},t\rangle|^2]\leq 4N\sqrt{N}s
\eeq
We may now integrate this inequality using the initial conditions (\ref{ci}):
\beq
\sum_{m,m'} [1-|\langle\psi_m,t|\psi_{m'},t\rangle|^2] \leq 4 N\sqrt{N} \int_0^T s(t) dt
\eeq
Finally, using condition (\ref{condition}) and $0\leq s(t)\leq1$, we find
\beq
T \geq \frac{\varepsilon}{4} \frac{N-1}{\sqrt{N}}
\eeq
or, with $N>>1$,
\beq
T \geq \frac{\varepsilon}{4} \sqrt{N}
\eeq
We conclude that, in order to be able 
to distinguish between states corresponding to different solutions
($\varepsilon > 0$), 
the computation must last a minimum time of order $\sqrt{N}$, 
which is what we found in Eq. (\ref{complexity}). 
Our choice of $s(t)$ is thus optimal.

\end{multicols}


\begin{references}

\bibitem{farhi1} E. Farhi, J. Goldstone, S. Gutmann, and M. Sipser, ``Quantum
Computation by Adiabatic Evolution'', quant-ph/0001106.

\bibitem{shor} P. W. Shor, in {\it Proceedings of the 35th Annual
Symposium on the Foundations of Computer Science}, 1994, Los Alamitos, 
California, edited by S. Goldwasser (IEEE Computer Society Press, New York,
1994), pp. 124-134.
 
\bibitem{grover} L. K. Grover, ``Quantum mechanics helps in searching for a
needle in a haystack'', Phys. Rev. Lett. {\bf 79}, 325 (1997).

\bibitem{farhi2} E. Farhi and  S. Gutmann, ``An Analog Analogue of a Digital
Quantum Computation'', quant-ph/9612026; Phys. Rev. A {\bf 57}, 2403
(1998).

\bibitem{farhi3} E. Farhi, J. Goldstone, S. Gutmann, J. Lapan, A. Lundgren and D. Preda,
``A Quantum Adiabatic Evolution Algorithm Applied to an NP-complete Problem'', quant-ph/0104129.

\bibitem{cerf} N. J. Cerf, L. K. Grover, and C. P. Williams,
``Nested quantum search and structured problems'', quant-ph/9806078;
Phys. Rev. A {\bf 61}, 032303 (2000).

\bibitem{bransden} B. H. Bransden and C. J. Joachain, {\it Quantum Mechanics},
(Pearson Education, 2000).


\end{references}
\end{document}